\documentclass[aps,twocolumn,pra,superscriptaddress,amsmath,showpacs,tightenlines]{revtex4}
\usepackage{graphicx}

\usepackage{times}
\usepackage{tgtermes}

\begin{document}
\title{Strategy for implementing stabilizer-based codes on solid-state qubits}

\author{Tetsufumi Tanamoto}
\affiliation{Corporate Research and Development Center, Toshiba Corporation,
Saiwai-ku, Kawasaki 212-8582, Japan}

\author{Vladimir M. Stojanovi\'c}
\affiliation{Department of Physics, University of Basel,
  Klingelbergstrasse 82, CH-4056 Basel, Switzerland}

\author{Christoph Bruder}
\affiliation{Department of Physics, University of Basel,
  Klingelbergstrasse 82, CH-4056 Basel, Switzerland}

\author{Daniel Becker}
\affiliation{Department of Physics, University of Basel,
  Klingelbergstrasse 82, CH-4056 Basel, Switzerland}
\date{\today}

\begin{abstract}

We present a method for implementing stabilizer-based codes with
encoding schemes of the operator quantum error correction paradigm,
e.g., the ``standard'' five-qubit and CSS codes, on solid-state qubits
with Ising or $XY$-type interactions. Using pulse
sequences, we show how to dynamically generate the effective dynamics of the
stabilizer Hamiltonian, the sum of an appropriate set of stabilizer
operators for a given code.  Within this approach, the encoded states
(ground states of the stabilizer Hamiltonian) can be prepared without
measurements and preserved against both the time evolution governed by
the original qubit Hamiltonian, and errors caused
by local sources.
\end{abstract}
%
\maketitle

\section{Introduction}

A variety of quantum error-correcting codes (QECCs) have been widely
investigated aiming at a robust computing system similar to the
classical digital
computer~\cite{Gottesman,Nielsen,Kitaev, Dennis,Kribs,Poulin,Bacon,Bombin,Bravyi,Milman,Knill1,Knill2,Reed,Moussa,korotkov2012}.
In particular, codes based on the stabilizer formalism constitute an
important class of QECCs.  This formalism has proven useful not only
for the standard codes~\cite{Gottesman,Nielsen}, but also for the
subsystem code~\cite{Kribs,Poulin,Bacon},
topological~\cite{Kitaev, Dennis, Bombin,wootton_loss_2012}, 
and Majorana codes~\cite{Bravyi}. On
the experimental side, Knill {\it et al.} demonstrated its usefulness
in the NMR domain~\cite{Knill1,Knill2}.  Stabilizer-based QECCs in
systems with always-on coupling have recently attracted a great deal
of interest~\cite{Nigg_Girvin:12,De_Pryadko:12}.

Stabilizer operators $G_j$ ($j=1,\ldots,l$) are mutually commuting
operators given by products of multiple Pauli matrices $X_i$, $Y_i$,
and $Z_i$ ($i=1,\ldots,n$)~\cite{Nielsen}. Conventionally, logical
qubit states are encoded through measurements into a joint,
$2^l$-dimensional, eigenspace $\mathcal{H}_S$ of these operators. For
$l$ stabilizer operators and $n$ physical qubits, a maximum number of
$k = n - l$ logical qubits can be encoded into $\mathcal{H}_S$, while
$k < n - l$ in case of subsystem encoding.  Since the ground states of
the stabilizer Hamiltonian $H_{\rm stab} :=-\sum_{j=1}^l G_j$ are
joint eigenstates of all stabilizer operators, its ground-state
manifold can play the role of $\mathcal{H}_S$.

It is important to note that stabilizer operators of many
error-correction codes, e.g., the surface code~\cite{Kitaev} or color
code~\cite{Bombin}, are given by products of more than two Pauli
matrices.  Therefore the corresponding stabilizer Hamiltonians cannot
be directly implemented in natural solid-state qubit systems, where
the interactions between qubits are of two-body
type~\cite{tana,jacobs2012}.

In this work, we demonstrate how to prepare ground states of
$H_{\rm stab}$ as encoded states and preserve them by inducing the effective
dynamics of this Hamiltonian using sequences of pulses in the form of
single-qubit rotations.  Being based on single-qubit rotations only, our method
works for an always-on physical (qubit) Hamiltonian with two-qubit
interactions, i.e., it does not require switching on and off any of its parts
(single-qubit or interaction). That local manipulation schemes are
in general sufficient to induce arbitrary Hamiltonian dynamics was shown by
Bennet \emph{et al.}~\cite{bennet2002}. Benjamin and Bose~\cite{benjamin2003} proposed
a particular implementation based on single-qubit rotations to perform quantum
computations on a one-dimensional system of bare qubits with always-on
Heisenberg interactions.

The distinguishing feature of our method is that it allows the preparation of
QECC encoded states without measurements, thus avoiding
measurement-induced decoherence. The method can be used not only for standard
codes (i.e., five-qubit and CSS codes) but also for the extended class of codes
with encoding schemes within the general operator quantum error correction
framework~\cite{Kribs,Poulin}.  Even in the presence of inevitable pulse
(rotation angle) errors the ground-state fidelity scales favorably with the
system size. 

Our scheme provides an essential ingredient for the implementation
of stable solid-state quantum memories. This, in turn, facilitates the
realization of quantum gates~\cite{De_Pryadko:12} within the limitations imposed
by the size and coherence time of a system. In fact, our approach even allows us to
directly realize arbitrary single- and multiqubit gates on error-correcting
codewords using only single-qubit rotations. Although the scheme requires a
rather large number of pulses (rotations), its feasibility can be anticipated
based on the recent progress in qubit-manipulation
techniques~\cite{BulutaEtAl:11}.

The paper is organized as follows. After the introduction of the
scheme to induce the dynamics of a stabilizer operator starting from a simple initial
Hamiltonian in Sec.~\ref{sec:DynGen}, we explain the extraction of suitable
initial Hamiltonians and two-qubit gates from typical solid-state qubit
Hamiltonians using only single-qubit rotations in Sec.~\ref{sec:ExtractFromH}.
The whole procedure is applied to the examples of the five-qubit, Steane, and
Kitaev's surface code in Sec.~\ref{sec:StdCodes}, illustrating the versatility
and generality of our method. This is followed by an illustration of how to use
pulses both to prepare codewords without measurements and apply gate operations
on logical qubits in Sec.~\ref{sec:PrepAndGate}. A discussion of the robustness of the scheme
against pulse errors is provided in Sec.~\ref{sec:Rubustness}. Finally, in Sec.
\ref{sec:Conclusions}, we present our conclusions.


\begin{table*}
\caption{%
\label{tab:5QB}
Stabilizer operators for the five-qubit code~\cite{Gottesman} as realized with
$XY$-type interactions. Starting from $H_{\rm ini}$, the generating sequence
will realize the desired stabilizer operator. Each appearance of $H^{i,i+1}_{XY}$
indicates an application of the transformation $H \to\; e^{-i \tau_{\rm op}
H_{\rm op}} H e^{i \tau_{\rm op} H_{\rm op}}$, while each instance of
$(\pi/2)_i^x$ denotes a $\pi/2$ rotation about the $x$ axis. The rightmost
column shows the time required to generate each stabilizer operator; $\tau_{\rm
rot}$ is the time needed to perform a single-qubit rotation.
}
\begin{tabular}{l|lllll|l|l|l}
\hline\hline
& & & & & & $H_{\rm ini}$ & generation sequence
&time for generation \\
\hline
$G_1$   & $X$ & $Z$ & $Z$ & $X$ & $I$ 
& $\Omega_2 X_2$ 
& $H_{XY}^{23}\rightarrow (\pi/2)_2^x\rightarrow H_{XY}^{12}+H_{XY}^{34}$ 
& $\tau_{\rm ini}+24\tau_{\rm rot}+4\tau_{\rm op}$ \\

$G_2$   & $I$ & $X$ & $Z$ & $Z$ & $X$ 
& $\Omega_3 X_3$ 
& $H_{XY}^{34}\rightarrow (\pi/2)_3^x \rightarrow H_{XY}^{23}+H_{XY}^{45}$
& $\tau_{\rm ini}+24\tau_{\rm rot}+4\tau_{\rm op}$ \\

$G_3$   & $X$ & $I$ & $X$ & $Z$ & $Z$ 
&$\Omega_2 X_2$ 
&$G_1\rightarrow H_{XY}^{45}\rightarrow (\pi/2)_2^x(\pi/2)_5^x 
\rightarrow H_{XY}^{23}$
&$\tau_{\rm ini}+43\tau_{\rm rot}+8\tau_{\rm op}$ \\

$G_4$   & $Z$ & $X$ & $I$ & $X$ & $Z$ 
&$\Omega_2 X_2$ 
&$G_1\rightarrow H_{XY}^{45}\rightarrow (\pi/2)_3^x(\pi/2)_5^x 
\rightarrow H_{XY}^{23}
\rightarrow (\pi/2)_1^y(\pi/2)_4^y$
&$\tau_{\rm ini}+45\tau_{\rm rot}+8\tau_{\rm op}$ \\ 

\hline\hline
\end{tabular}
\end{table*}


\section{Dynamical generation of stabilizer operators}\label{sec:DynGen}

As a first step, we show how a stabilizer operator $G_j$ can be dynamically
generated from a simple initial Hamiltonian $H_{\rm ini}\propto X_i$, $Y_i$, or
$Z_i$. The time evolution corresponding to the generation process is illustrated
with the schematic notation $\rho(0) \stackrel{t H}{\longrightarrow } \rho(t)$,
where $\rho(t) = \exp(-iHt) \rho(0) \exp(iHt)$ is the density matrix for a
time-independent Hamiltonian $H$, or for an effective $H$ in the sense of
average-Hamiltonian theory~\cite{Ernst}. After the application of mutually inverse, unitary
operations according to $\rho(0) \stackrel{\tau_{\rm op} H_{\rm op}
}{\longrightarrow } \ \ \stackrel{ \tau_{\rm ini} H_{\rm ini} }{\longrightarrow
} \ \ \stackrel{ -\tau_{\rm op} H_{\rm op} }{\longrightarrow } \rho(\tau_{\rm
ini}+2\tau_{\rm op})$, the system has evolved as if propagated by the effective
Hamiltonian $\exp(-i \tau_{\rm op} H_{\rm op}) H_{\rm ini} \exp(i \tau_{\rm op}
H_{\rm op})$ for a time $\tau_{\rm ini}$~\cite{stab}. To build the stabilizer
operator $G_j$ from $H_{\rm ini}$, we need two elementary transformations: one
that rotates arbitrary single-qubit terms through an angle of $\pi/2$ and
another one that increases the order of Pauli-matrix terms by $1$. If $H_{\rm
op}$ is the generator of a single-qubit rotation, say $-J X_i$, such a sequence
dynamically generates the time evolution of $H_{\rm ini}$ rotated about the
$x$ axis through angle $2 J \tau_{\rm op}$. Higher-order products of Pauli
matrices can be generated using the following transformations~\cite{stab}:
\begin{eqnarray}
e^{-it H_{XY}^{i,i+1}} X_i e^{it H_{XY}^{i,i+1}} &=& c_\theta X_i 
-s_\theta Z_i Y_{i+1}\:,\nonumber\\
e^{-it H_{XY}^{i,i+1}} Y_i e^{it H_{XY}^{i,i+1}} &=& c_\theta Y_i 
+s_\theta Z_i X_{i+1}\:,\nonumber\\
e^{-it H_{XY}^{i,i+1}} Z_i e^{it H_{XY}^{i,i+1}} &=& c^2_\theta Z_i 
+s^2_\theta Z_{i+1}\nonumber\\
&+&c_\theta s_\theta (X_iY_{i+1}-Y_iX_{i+1})\:,
\end{eqnarray}
where $H_{\rm XY}=\sum_{i} H_{XY}^{i,i+1}\equiv
J\sum_{i}(X_iX_{i+1}+Y_iY_{i+1})$ is the (two-body) $XY$ interaction and
$c_\theta \equiv \cos (2\theta)$ and $s_\theta \equiv \sin (2\theta)$. For
$\theta=Jt=\pi/4$, these transformations increase the order of the Pauli-matrix
terms as $X_i \rightarrow - Z_i Y_{i+1}$ and $Y_i \rightarrow Z_i X_{i+1}$.
Similarly, one obtains $Z_i \rightarrow Z_{i+1}$. Analogous relations hold for
the Ising interaction given by $H_{\rm Ising}=J\sum_{i}Z_i Z_{i+1}$.

With $\tau_{\rm op} = \pi/(4 J)$ and a properly constructed, nested sequence of
operations $H_{\rm op}$, framing a period of propagation with $H_{\rm
ini}$ and duration $\tau_{\rm ini}$, we can therefore induce the dynamics of arbitrary
stabilizer operators $G_j$. Table I shows such sequences for the case of the five-qubit code.
The time $\tau_{\rm ini}$ has to be chosen such that the entire process can be
carried out in a time interval sufficiently shorter than the coherence time.

\section{Extracting $H_{\rm ini}$ and $H_{\rm op}$ from a qubit Hamiltonian}\label{sec:ExtractFromH}

The key step in dynamically generating the stabilizer operators is extracting a
single-qubit part or a pure two-body interaction part from a qubit system with
Hamiltonian $H=H_0+H_{\rm XY}$, where $H_0=\sum_i H_{0i}=\sum_i(\Omega_i
X_{i}+\varepsilon_i Z_{i})$ is a single-qubit part. Note that instead of the
$XY$ Hamiltonian we could also use the Ising Hamiltonian.

This process is carried out using the Baker-Campbell-Hausdorff (BCH) formula
\cite{Ernst}. For simplicity, we explain this procedure for $\varepsilon_i = 0$
in $H_0$, where only rotations about the $z$ axis will be needed, and set
$\Omega_i = \Omega$. In the general case, the procedure requires a slightly more
complex pulse sequence.

A part $H_a$ can be extracted from $H_0$ by applying a single appropriate $\pi$
pulse, if that pulse transforms $H_0$ to $H_a - H_b$, where $H_b = H_0 - H_a$
consists of the unwanted terms. For $2n$ alternating periods of propagation with
$A=i \tau (H_a+H_b)$ and $B=i \tau (H_a-H_b)$, the BCH formula yields
\begin{equation}
(e^Ae^B)^n \approx \exp( i 2n\tau H_a + n\tau^2[H_a,H_b])
\label{AB}
\end{equation} 
where the duration of the pulse sequence is $2 n \tau$. Thus, as long as
$\tau||H_b|| \ll 1$, where $||A||=[\mathrm{Tr}(A^\dagger A)/d]^{1/2}$ is the
standard operator norm in a Hilbert space of dimension $d$, we can neglect the
second term.  As the number $n$ of repetitions increases, this approximation
becomes progressively better.

In order to extract a single-qubit (local) part of the system
Hamiltonian, relation (\ref{AB}) has to be applied twice, leading to (case $n=1$)
\begin{eqnarray}\label{ABB'A'}
\lefteqn{ e^Ae^B  e^{B'}e^{A'} \approx \exp( 2h_a + [h_b,h_a]) \exp( 2h_a' - [h_b',h_a'])} \nonumber \\
      & \approx& \exp\{ 2(h_a+h_a') + [h_b,h_a]- [h_b',h_a'] +4 [h_a, h_a'] \}, 
\label{ABAB}
\end{eqnarray}
where $h_{a/b}^{(\prime)} := i \tau H_{a/b}^{(\prime)}$, $A'=h_a'+h_b'$ and
$B'=h_a'-h_b'$. Consider extracting a single-qubit Hamiltonian $X_2$ for a
one-dimensional five-qubit array. With $h_i=i \tau \Omega X_i$ (single-qubit
Hamiltonian with $\epsilon_i=0$) and $h_{ij}=i \tau H_{XY}^{ij}$, we set
\begin{eqnarray}
h_a &=& h_{2}+h_{34}+h_{45},   \nonumber \\
h_b &=& h_{1}+h_{3}+h_{4}+h_{5}+h_{12}+h_{23},  \nonumber \\
h_a'&=& h_{2}-h_{34}-h_{45},   \nonumber \\
h_b'&=& h_{1}-h_{3}+h_{4}-h_{5}+h_{12}-h_{23}.  
\end{eqnarray}

The sequence of operators describing the time evolution on the
left-hand side of Eq.~\eqref{ABB'A'} is obtained in the following manner:
By applying a $\pi$ pulse to qubits 1, 3, 4, and 5 one transforms $A$ into
$B$ and $B'$ into $A'$, while $B$ is transformed into $B'$ 
by a $\pi$ pulse applied to qubits 3 and 5.
This leads to $H_{\rm ini}=H_{\rm ini}^{(0)}+H_{\rm ini}^{(1)}$, where
$H_{\rm ini}^{(0)}=\Omega X_2$ is the desired initial Hamiltonian and
$H_{\rm ini}^{(1)} = [\Omega \tau/2] (J_{32} Y_3Z_2 -J_{34} Y_3Z_4
-J_{45} Y_5Z_4)$ is an unwanted perturbation term. This term scales
like $J_{ij}\tau \ll 1$ and hence can be reduced by shortening the
duration of the pulse sequence. 

Similarly, the operator $e^{-it H_{XY}^{12}}$ is obtained by 
extracting $H_{XY}^{12}$ from the system Hamiltonian using 
\begin{eqnarray}
h_a \!\!\!&=& \!\! h_{23}+h_{1}+h_{4},  \nonumber \\
h_b \!\!\!&=& \!\! h_{12}+h_{34}+h_{45}+h_{2}+h_{3}+h_{5},  \nonumber \\
h_a' \!\!\!&=& \!\! h_{23}-h_{1}-h_{4},  \nonumber \\
h_b' \!\!\!&=& \!\! -h_{12}-h_{34}-h_{45}+h_{2}+h_{3}+h_{5}. 
\end{eqnarray}
The perturbation terms 
can be neglected for $J/\Omega \ll 1$.

In the following, we apply our scheme to the five-qubit code and Steane's
seven-qubit code (the smallest single-error correcting CSS code)~\cite{Gottesman}, as well
as the surface code~\cite{Kitaev}.

\section{Realization of five-qubit, Steane, and Kitaev's surface codes} \label{sec:StdCodes}

The generation processes of the four stabilizer operators $G_j
(j=1,\ldots,4)$ of the five-qubit code~\cite{Gottesman} 
are shown in Table I. For example, starting from the initial 
Hamiltonian $H_{\rm ini}=\Omega_2 X_2$, 
the stabilizer operator $G_1$ of the five-qubit code is realized
through the sequence
\begin{eqnarray}
e^{-i\tau_{\rm op} H_{XY}^{23}}X_2e^{i\tau_{\rm op}
  H_{XY}^{23}}&\rightarrow&-Z_2 Y_3 \nonumber\\
-e^{-i(\pi/4) X_2 } Z_2 Y_3e^{i(\pi/4) X_2 } &\rightarrow& Y_2 Y_3\\
e^{-i\tau_{\rm op} (H_{XY}^{12}+H_{XY}^{34})}Y_2 Y_3
       e^{i\tau_{\rm op} (H_{XY}^{12}+H_{XY}^{34})}&\rightarrow& X_1Z_2
       Z_3X_4\:. \nonumber
\end{eqnarray}
The minimal time required for this process is 
$\tau_{\rm ini}+ 24 \tau_{\rm rot} + 4 \tau_{\rm op}$.  
The effective dynamics of $H_{\rm stab}=-\sum_{l=1}^4 G_l$ is induced
by subsequent generation of the four stabilizer operators.

We would now like to address the feasibility of this scheme in a
typical superconducting qubit system.
For two superconducting qubits in a circuit-QED setup 
the resulting effective interqubit interaction is also of $XY$
type~\cite{Blais++:04,toffoli}.  For instance, for $g/\Delta=0.1$, $g/(2\pi)=200$
MHz, $\Delta/(2\pi)=2$ GHz, where $g$ is the Jaynes-Cummings coupling
constant and $\Delta$ is the detuning between the resonator frequency and
the qubit splitting, we have $J/(2\pi)=20$ MHz. Assuming $\tau_{\rm rot}\sim
1$ ns~\cite{chow2010}, we obtain a  minimal total time of 
$\tau_{\rm 5code}^{\rm min}=24\tau_{\rm op} +136 \tau_{\rm rot}
\approx 300$ ns, 
which is significantly shorter than $T_2 \sim 20$ $\mu$s 
reported in~\cite{Paik}.

\begin{table*}
\caption{%
\label{tab:CSS}
Stabilizer operators for the Steane code~\cite{Gottesman} as
realized with $XY$-type interactions.  The operators $G_4$, $G_5$, and
$G_6$ are obtained by replacing $X$ with $Z$ in $G_1$, $G_2$, and
$G_3$, respectively.  The rightmost column shows the time required to
generate each stabilizer operator; $\tau_{\rm rot}$ is the time needed
to perform a single-qubit rotation.
}%
\begin{tabular}{l|lllllll|l|p{8.5cm}|l}
\hline\hline
& & & & & & & & $H_{\rm ini}$ &  generation sequence
&time for generation \\
\hline
$G_1$ & $X$ & $X$ & $X$ & $X$ & $I$ & $I$ & $I$ 
&$\Omega_2 X_2$
&$H_{XY}^{23}\rightarrow (\pi/2)_2^x\rightarrow H_{XY}^{12}
+H_{XY}^{34}\;\rightarrow (\pi/2)_2^x (\pi/2)_3^x$ 
&$\tau_{\rm ini}+26\tau_{\rm rot}+4\tau_{\rm op}$\\
$G_2$ & $X$ & $X$ & $I$ & $I$ & $X$ & $X$ & $I$ 
&$-\Omega_3 X_3$
&$ H_{XY}^{34} \rightarrow  (\pi/2)_3^y \rightarrow 
 H_{XY}^{23}+H_{XY}^{45} \rightarrow H_{XY}^{12}+H_{XY}^{56} \rightarrow  
 (\pi/2)_3^x(\pi/2)_4^x(\pi/2)_6^x \rightarrow  H_{XY}^{23}+H_{XY}^{45} 
\rightarrow  (\pi/2)_6^y $
&$\tau_{\rm ini}+45\tau_{\rm rot}+8\tau_{\rm op}$\\
$G_3$ & $X$ & $I$ & $X$ & $I$ & $X$ & $I$ & $X$ 
&$-\Omega_3 X_3$
&$ H_{XY}^{34} \rightarrow 
(\pi/2)_3^x \rightarrow 
 H_{XY}^{23}+H_{XY}^{45} \rightarrow 
H_{XY}^{12}+H_{XY}^{56} \rightarrow 
 H_{XY}^{67} \rightarrow 
 (\pi/2)_1^x(\pi/2)_2^x(\pi/2)_4^x(\pi/2)_6^x  \rightarrow 
 H_{XY}^{12}+H_{XY}^{34}+H_{XY}^{56}$
&$\tau_{\rm ini}+51\tau_{\rm rot}+10\tau_{\rm op}$\\
\hline\hline
\end{tabular}
\end{table*}

Table II shows how to generate the Steane code.
The stabilizer operators $G_4$, $G_5$, $G_6$ are obtained by 
$e^{-\pi \sum_i Y_i/4 }(G_1+G_2+G_3) e^{\pi \sum_i Y_i/4 }$.
Thus, the minimal total time is 
$\tau_{\rm Steane}^{\rm min} 
=44\tau_{\rm op}+246 \tau_{\rm rot}\approx 600$ ns,
i.e., again much shorter than the $T_2$ given in Ref.~\cite{Paik}.  

\begin{figure}
\includegraphics[width=8cm,clip=true]{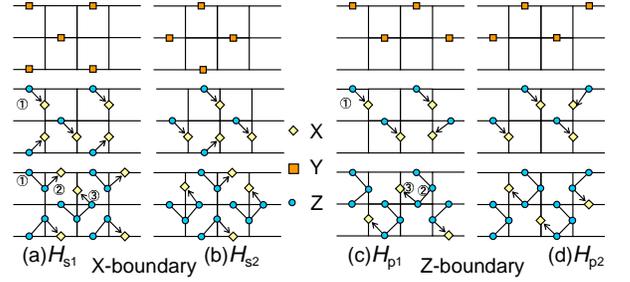}
\caption{(Color online) Four types of surface-code generation on the $2\times 3$
  lattice in Ref.~\cite{Kitaev} starting from a single-qubit
  Hamiltonian that includes $Y$ operators. All operators in (a) and (b)
  are transformed to $X$ and those in (c) and (d) to $Z$. Subsequently,
  the four types of stabilizer operators are combined into the topological
  Hamiltonian in the manner discussed in the text.}
\label{topo}
\end{figure}

For the realization of Kitaev's surface code, we need to generate four types of stabilizer operators.
Qubits are placed at the edges of the
square lattice; see Fig.~\ref{topo}. 
Stabilizer operators $H_s=\Pi_{j\in {\rm star(s)}}
X_j$ are assigned to each vertex $s$, and $H_p=\Pi_{j\in {\rm boundary} (p)}
Z_j$ to each face $p$. Using the relations $X_i\rightarrow
-Z_iY_{i+1}$ and $Y_i\rightarrow Z_iX_{i+1}$, we can form products of
nearest-neighbor operators such as $Y_1 \rightarrow Z_1X_2$
$\rightarrow -Z_1Z_2Y_3\rightarrow Z_1Z_2Z_3X_4$.  In generating
adjacent stabilizer operators, care should be taken to avoid mixing them. This
can be achieved by decompositions like $H_s=H_{s_1}+H_{s_2}$ 
[see Figs.~\ref{topo}(a) and \ref{topo}(b)] and $H_p=H_{p_1}+H_{p_2}$
[Figs.~\ref{topo}(c) and \ref{topo}(d)].  
The four elements can then be combined into the total surface-code Hamiltonian.

\section{Preparation of encoded states  and gate operations} \label{sec:PrepAndGate}

Our approach also allows us to prepare encoded states (or codewords) 
of general stabilizer-based codes without performing
measurements on the system and to implement arbitrary single- and multiqubit gate operations.

We show this in detail for the standard
codes, which encode $k$ logical qubits into a subspace of dimension
$2^k$. However, this procedure also works for
subsystem encoding provided suitable stabilizer operators are added. 
For any given code,
only those $G_j$ with $1 \le j \le m$ and $m \le n - k$ that contain
$X$ or $Y$ operators are needed for the preparation:
\begin{eqnarray}\label{eqn:CodeGen}
|\bar{c}_1\cdots \bar{c}_k\rangle &=& (1+G_1)\cdots
(1+G_{m})\bar{X}^{c_1}_1
\cdots\bar{X}^{c_k}_k |0\cdots 0\rangle \nonumber \\
&= & \prod_{i=1}^{k} \bar{X}^{c_i}_i \prod_{j=1}^{m} 
\exp\left( i \frac{\pi}{4}  \tilde{G}_j^{a_j} \right) |0...0\rangle\:,
\label{gen}
\end{eqnarray}
where $c_i=0,1$ and operators $\bar{X}_i$ act in the logical state space
$\{|\bar{0}\rangle_i$, $|\bar{1}\rangle_i\}$.  Here, $\tilde{G}_j^{a_j}$ denotes 
a modified stabilizer operator obtained from $G_j$
by replacing the $X$ operator acting on qubit $ a_j$ by a $Y$
operator, or vice versa. This is done in order to match the effect of an \emph{individual} 
factor $\exp [i(\pi/4) \tilde{G}_j^{a_j} ]$ with
the action of the projector $(1+G_j)$ when qubit $ a_j$ is in state
$|0\rangle$. To fulfill Eq.~\eqref{eqn:CodeGen} for all $1 \le j \le m$ \emph{simultaneously}, all
the $a_j$ have to be different and the modified stabilizers have to be generated
in an order such that prior to $\tilde{G}_j^{a_j}$ none of the $\tilde{G}_k^{a_k}$
with $k < j$ have acted on qubit $a_j$ with an $X$ or $Y$.

By implementing the second row of Eq.~\eqref{eqn:CodeGen}, the quantum
information is encoded into the logical qubit \emph{after} the basis state $| \bar{0} \rangle$
is generated by applying appropriate logical gate operations
(see below). It is also possible, however, to start from an arbitrary
(potentially unknown) qubit state that is encoded into a $2^k$-dimensional
subspace of physical qubits $b_l \ne a_j$ for all $j$ with $1 \le l \le n - m$.
For $m < n - k$ this subspace has to be a simultaneous eigenspace of the
stabilizer operators $G_{m + 1}, \ldots, G_{n - k}$, which contain only $Z$ operators
[as the $m$ stabilizer operators that involve $X$ and $Y$ are used for state preparation,
in accordance with Eq.~\eqref{eqn:CodeGen}].
This second approach is particularly useful for codes with $k = 1$ and $m = n -
k$, like the five-qubit code. In that case, for any choice of qubits $a_j$ the
generation of the $\tilde{G}_j^{a_j}$ alone directly encodes the state of the
single physical qubit $b \ne a_j$ into the ground-state manifold of the
stabilizer Hamiltonian.

We illustrate the encoding procedure on the example of a
three-qubit code whose stabilizer operators are $X_1X_2$ and $X_2X_3$.
This is realized in a three-qubit system with Ising interactions. 
The stabilizer Hamiltonian $J(X_1X_2+X_2X_3)$ is
obtained by removing the single-qubit part $H_0$ of the original
Hamiltonian using $\pi$ pulses.  
Its ground states can be written as $|\bar{c}\rangle = |c\rangle \otimes (
|00\rangle + |11\rangle ) + |1 - c\rangle \otimes ( |01\rangle + |10\rangle )$
with $c = 0,1$. Thus, an arbitrary logical single qubit state $| \bar{\alpha}
\rangle := \cos (\alpha) | \bar{0} \rangle + \sin(\alpha) | \bar{1} \rangle$ can
be obtained via $| \bar{\alpha} \rangle = \frac{1}{2}(1+X_2X_3)(1+X_1X_2)|\alpha
0 0 \rangle=\exp[i(\pi/4)X_2Y_3]\exp[i(\pi/4)X_1Y_2]|\alpha 00 \rangle$. Note
that since the choice of modified stabilizers is not unique, we could just as
well start from a state with the information initially encoded in qubit 2 or 3. 

For the five-qubit code, we  can choose $ \tilde{G}_1^1 = Y_1Z_2Z_3X_4$,
$ \tilde{G}_2^5 = X_2Z_3Z_4Y_5$, $ \tilde{G}_3^3 = X_1Y_3Z_4Z_5$,
$ \tilde{G}_4^2=Z_1 Y_2 X_4 Z_5$,  where the multiplication in
Eq.~(\ref{gen}) is carried out in the following order: $\exp[ i
  (\pi/4) \tilde{G}_2 ]$ $\exp[i (\pi/4) \tilde{G}_4 ]$ $\exp[ i
  (\pi/4) \tilde{G}_3 ]$ $\exp[i (\pi/4) \tilde{G}_1 ]$.
This choice of modified stabilizers encodes the state of qubit
4 into the corresponding codeword state.
In the case of the Steane code, $ \tilde{G}_1^4 = X_1X_2X_3Y_4$,
$ \tilde{G}_2^6=X_1X_2X_5Y_6$, $ \tilde{G}_3^7 = X_1X_3X_5Y_7$. Note that here only
three out of six stabilizer operators are needed for the preparation of an
encoded state.

Gate operations on logical states, like rotations, phase gates, etc., can be
realized by dynamically generating the generators of the gates for an
appropriate time. For example, the Pauli operator $\bar{X}$ for the five-qubit
code is given by $\bar{X} = X_1 \cdots X_5 = \exp (i \pi X_1 \cdots X_5 / 2)$.
Hence, by dynamically generating the average Hamiltonian $-\Omega X_1 \cdots
X_5$ within a time $t = \phi / (2 \Omega)$, we can perform a rotation through angle
$\phi$ about the $x$ axis on state $|\bar{\alpha} \rangle$. For two five-qubit
codes implemented on physical qubits 1 to 10, the two-qubit controlled phase
gate is applied by generating $-\Omega Z_1 \cdots Z_{10}$ within a time $t = \pi/(4
\Omega)$. The generalization to arbitrary gate operations and codes is
straightforward.

\section{Robustness against pulse errors} \label{sec:Rubustness}

Since the codeword states are encoded in the twofold-degenerate ground-state
manifold $|\bar{0}\rangle$ and $|\bar{1}\rangle$ of $H_{\rm stab}$, the
robustness of this method is limited by the rate of leakage out of this
manifold.  In principle, precise estimates of the leakage due to the thermal
environment could be obtained by studying the stability of the ground state to
various perturbations as in Ref.~\cite{Bravyi2}. However, energy
nonconserving single-qubit errors---often a prevalent kind of
error created by a thermal bath---are exponentially suppressed
for temperatures that are small compared to the Zeeman-splitting $\Omega$.
Hence, besides local imperfections and noise sources, unavoidable
pulse errors are likely to be the predominant cause of leakage, at
low temperatures.

To estimate this effect, we consider pulse errors that can be modeled
by randomly distributed, unbiased, and uncorrelated deviations
$\delta\theta$ with $\sigma_\theta = \sqrt{\langle \delta \theta^2 \rangle}$
from the ideal angle of $\pi / 2$. The leakage can then be estimated
by looking at the average of the ground-state fidelity $F (t) =
\lvert\langle \bar{0}\vert U_{\rm P} (t) \vert \bar{0}\rangle \rvert^2
$, where $U_{\rm P} (t)$ is the time evolution operator with imperfect
pulses. This average is approximately given by $\langle F (t) \rangle
\approx 1 - N_{\rm P} \sigma_\theta^2 t / (8 \mathcal{T})$, where $N_{\rm P}$
is the number of pulses in the sequence to generate $H_{\rm stab}$,
and $\mathcal{T}$ is its duration.  The number $N_{\rm P}$ of pulses is
given by the number of rotations needed to generate all stabilizers
of a given code (for the five-qubit and Steane code, see
Tables~\ref{tab:5QB} and \ref{tab:CSS}, respectively).

\section{Conclusions} \label{sec:Conclusions}

In conclusion, we have demonstrated measurement-free preparation of
encoded states in the stabilizer-based codes described by the operator
quantum error correction paradigm.  The scheme is based on pulse
sequences applied to solid-state qubit Hamiltonians with two-body
interactions of $XY$ or Ising type.  We have estimated the intrinsic
robustness of our scheme against pulse imperfections.
Depending on the required operation time needed for scalable quantum
computations using a particular (solid-state) qubit implementation, this allows us to
determine an upper limit for the magnitude of pulse errors. In addition to being
the first step of realizing a large class of stabilizer codes in solid-state
systems at all, the dynamic generation of QECC Hamiltonian dynamics also
provides protection against certain classes of local errors such as impurities.

Further steps towards a stable quantum memory would require to protect the code
against thermal fluctuations, which could be achieved, e.g., by a coupling to
appropriate, nonlocal external fields~\cite{Chesi2010, Pedrocchi2012, Becker2013}.
Once implemented experimentally, our scheme will therefore pave the way for robust
quantum information processing.

We would like to thank Fabio Pedrocchi and Adrian Hutter for discussions.
This work was financially supported by the Swiss SNF and NCCR Quantum
Science and Technology.


\end{document}